\pdfoutput=1
\documentclass[manuscript,screen,nonacm]{acmart}
\AtBeginDocument{%
  }

\setcopyright{none}
\newcommand{\fillin}[1]{\textcolor{red}{[#1]}}

\usepackage{booktabs}
\usepackage{multirow}
\usepackage{tabularx}
\usepackage{array}
\usepackage{footnote}

\makeatletter
\def\@correspondingauthormark{\g@addto@macro\@currentauthors{%
    \advance\hfuzz by 5pt\relax\textsuperscript{*}\relax}}
\makeatother
\begin{document}

\title{Beyond Screen Time: Demonstrating the Value of App Activity Logs to Understand User Behavior Further}

\author{Ole Schmitt}
\email{ole.schmitt@student.hpi.de}
\correspondingauthor
\affiliation{%
  \institution{Hasso Plattner Institute}
  \city{Potsdam}
  \state{Brandenburg}
  \country{Germany}
}
\author{Pauline Gieseler}
\correspondingauthor
\email{pauline.gieseler@hpi.de}
\affiliation{%
  \institution{Hasso Plattner Institute}
  \city{Potsdam}
  \state{Brandenburg}
  \country{Germany}
}

\author{Frederik Riedel}
\affiliation{%
 \institution{riedel.wtf GmbH}
 \city{Berlin}
 \state{Berlin}
 \country{Germany}}

\author{Donatus Wolf}
\affiliation{%
  \institution{University of Potsdam}
  \city{Potsdam}
  \state{Brandenburg}
  \country{Germany}}

\author{Ariel Dora Stern}
\affiliation{%
  \institution{Hasso Plattner Institute}
  \city{Potsdam}
  \state{Brandenburg}
  \country{Germany}}

\author{Paul Schmiedmayer}
\affiliation{%
  \institution{Stanford University}
  \city{Stanford}
  \country{USA}}

\author{David J. Gr{\"u}ning}
\correspondingauthor
\affiliation{%
  \institution{Stanford University}
  \city{Stanford}
  \country{USA}}
\affiliation{%
  \institution{University of Cambridge}
  \city{Cambridge}
  \country{UK}}
\email{david.gruning@mrc-cbu.cam.ac.uk}
\affiliation{%
  \institution{Max-Planck Institute for Human Development}
  \city{Berlin}
  \country{Germany}}
\email{gruening@stanford.edu}

\renewcommand{\shortauthors}{Schmitt et al.}

\authorsaddresses{\textsuperscript{*}Corresponding authors: Ole Schmitt,
  ole.schmitt@student.hpi.de;
  Pauline Gieseler,
  pauline.gieseler@hpi.de;
  David J. Gr{\"u}ning,
  gruening@stanford.edu}.

\begin{abstract}
  Research on smartphone use remains fragmented, with studies differing in measures, methods, and platforms, limiting what we know about everyday behavior. We argue that event-level app activity logs should form the backbone of research on actual rather than recalled smartphone use. We conduct a secondary analysis of 4,571,252 app events from 1,972 participants across three longitudinal cohorts differing in age, country, recruitment, and platform, complemented by a published reference cohort. We reproduce established aggregate and micro-usage measures and examine temporal structure, application composition, transitions, and individual distinctiveness. Aggregate usage varies less than the organization of activity: adolescent use, for example, is structured around school schedules, while other cohorts show weaker within-day patterns. Application and transition patterns reveal behavioral differences obscured by screen time; an intervention reduced daily usage while increasing mean session duration. Activity traces also enabled 15–22\% top-1 participant re-identification. We discuss methodological, reproducibility, and privacy implications for HCI.
\end{abstract}



\maketitle

\section{Introduction}
\label{sec:introduction}
The topic of problematic technology use, specifically social media usage of adolescents entered the space of political discussions and policy making worldwide. Research in this space of mobile sensing and digital phenotyping has intensified investigating causal influences in mental health and smartphone use \cite{odgers_annual_2020,kim_real_2019}. 
Monitoring smartphone use and analysing total app usage, with active screen time and activity logs is stated to be the strongest behavioral biomarker for mobile sensing and digital phenotyping outperforming self-reported events and questionnaires \cite{torous_creating_2019,schroeder_digital_2026,li_effects_2024}. 
Existing research targets the behavioral monitoring to a great extend but leaving gaps between the different conducted studies. The most fundamental limitations of existing research is the lack of comparability of data collection, processing and analysis of monitoring pipelines and computing methods \cite{schroeder_digital_2026,hamilton_improving_2025}. 
With a look at these limitations, there are research gaps regarding the fine-grained temporal structure of screen time logs and the resulting individual behavioral signatures. Insights and methods for cross-app differences and within-person stability over long periods are missing.

This paper presents a secondary analysis of existing datasets containing objective smartphone-use metrics collected in studies using the \textit{one sec} app \cite{haliburton_longitudinal_2024, gruning_directing_2023}.
Using 4,571,252 logged smartphone app-opening events from 1,972 participants across three longitudinal studies, we examine app-opening behavior across cohorts that differ in age, country, recruitment strategy, and platform, and compare the findings with one additional sample to address the following research questions:

\textbf{RQ1} \textit{What patterns of app opening emerge from event-level logs, and can users be grouped according to the temporal structure of their activity rather than their total smartphone use?}

\textbf{RQ2} \textit{How many users exhibit unique usage patterns, and can an unlabeled activity trace be matched to the correct user, thereby enabling re-identification?}

\textbf{RQ3} \textit{Which limitations identified in previous studies using screen-time logging are consistently reproduced in the present datasets, and how can they be addressed methodologically?}

We further show that the micro-usage threshold \citet{ferreira_contextual_2014} shifts depending on whether it is derived via Jenks optimization or k-means, a choice earlier studies do not report. We argue that the comparability problem in this literature now sits in the conversion of event traces into features, and outline the reporting such analyses require.






\section{Related Work}
\label{sec:related_work}
Conceptually, our work builds on the research on psychological mechanisms underlying smartphone usage behavior and insights from longitudinal smartphone overuse studies from Human Computer Interaction research.
\subsection{Digital Practices of Adolescents and Their Mental Health}
\label{subsec:digital_practices_of_adolescents}
Even though there are already policy makers moving forward with regulating social media for adolescents they are moving ahead of evidence. 
In reviews of longitudinal it was concluded that causal relationships between social media use and metal health effects in adolescent were either too diminutive or confounded. This is rooted in study designs of small cohorts or even the definition of social media apps \cite{odgers_annual_2020}. There are contradictions remaining like excessive notification checking with a comparatively low total screen time \cite{hamilton_improving_2025}. These contradictions stretch over study designs when different study report a sum of micro-sessions of seconds can not be compared to a session in one app of the same length of the sum (e.g. 60 micro sessions of 60 seconds of different apps is not the same as 60 minutes in one app) \cite{bohmer_falling_2011,hansen_disrupting_2026,hamilton_improving_2025}.

\subsection{Why measuring is hard: Beyond Aggregate Usage Measures}
\label{subsec:beyond_aggregate_usage_measures}
Existing work on screen time analysis has surfaced the well-known comparability problem that conclusions about smartphone use shift with the dataset, population and logging method they are drawn from, making the need for replication exercises so necessary \cite{church_understanding_2015,kim_real_2019}.
The difficulty is not only that studies actually differ, but that these differences and the decisions by the researchers that led to them often are not well documented. That is, event-level traces must be converted into features before analysis, and such steps involve choices from the researchers that are rarely documented in enough detail to be reproduced, like temporal binning, the treatment of gaps as non-use or missing data, and the grouping of applications \cite{schroeder_digital_2026}. For illustration, \citet{hamilton_improving_2025} demonstrate this problem for app-grouping, finding that Google Play Store categories exclude YouTube (the single application used by every participant) and yield daily estimates roughly 30 minutes and at least 20 checks lower than researcher-coded categories.

\section{Methods}
\label{sec:methods}

\begin{table*}[]
\centering
\caption{Overview of the study populations and datasets used for replication}
\label{tab:demographics-studies}
\begin{tabular}{@{}lllll@{}}
\toprule
 & Adolescents Cohort (D) & Adolescents Cohort (G) & Alternative Cohort & Longitudinal Study (U)\\ \midrule
Population & $N = 243$ & $N = 262$ & $N = 1{,}467$ & $N = 19$ \\
Geography & Denmark & Germany & Worldwide & USA \\
Age groups in Years & 13--17 (M = 14.73) & 14 -- 25 (M = 19,5) & --- & 13--17 (M = 15.84) \\
Recruitment & Online panel & In-app & In-app & Clinical/community \\
Duration & 6 weeks & 6 weeks & 2 weeks & $\sim$31 days \\
Logging platform & \textit{one sec} & \textit{one sec} & \textit{one sec} & AWARE \\
Device platform & iOS & iOS  & iOS& Android \\
Event type & Open + close & Opening attempt & Opening attempt & App foreground \\
\midrule
Sessions / attempts & 353{,}796 & 200{,}889 & 3{,}662{,}771 & --- \\
\quad baseline & 159{,}449 & --- & --- & --- \\
\quad intervention & 194{,}347 & --- & --- & --- \\
\quad opened & --- & 128{,}145 & 1{,}303{,}689 & --- \\
\quad closed & --- & 30{,}936 & 1{,}700{,}077 & --- \\
\quad dismissed & --- & 41{,}808 & 659{,}005 & --- \\
Total events & 707{,}592 & 200{,}889 & 3{,}662{,}771 & 10{,}038\\
\midrule
Total duration & 25{,}574 h (2.92 y) & --- & --- & --- \\
\quad baseline & 10{,}709 h (1.22 y) & --- & --- & --- \\
\quad intervention & 14{,}865 h (1.70 y) & --- & --- & --- \\
Mean session duration & 4.34 min & --- & --- & --- \\
\quad baseline & 4.03 min & --- & --- & --- \\
\quad intervention & 4.59 min & --- & --- & --- \\ \bottomrule
\end{tabular}
\end{table*}

\subsection{Recruitment and Participants}
\label{subsec:participants}
\subsubsection{Adolescents Cohort (D)}
\label{subsubsec:adol-d}
Participants were 
adolescents aged 13-17 years using iPhones. 
Approximately 80\%  of 
adolescents use iPhones \cite{haliburton_longitudinal_2024}.
Participants were recruited through the online panel company Norstat A/S, who also managed consent, and incentive distribution.
Invitations were distributed to 1,981 municipal and public schools 
across Denmark. 
Participants aged 15 or lower required parental or guardian consent.
Of the 400 enrolled participants, 312 installed the app (\textit{one sec}) as well as the required shortcuts and automations successfully.
After excluding participants with technical issues or insufficient behavioral data 243 participants remained for analysis.
Technical errors led to the exclusion 43 participants, who either did not interact with any of the selected apps during the study period (22), lacked data during the intervention period (9), had no valid assignment to an intervention type (6), began the intervention phase to late (5), or did not open any selected app (1), yielding 269 participants. 

Due to the voluntary nature of participation, opting out during the course of data collection was possible. 
The exclusion of inactive participants (26), resulted in the final number of 243 participants included in the analysis and low attrition rate of 9.67\% evenly distributed among the intervention conditions. 
Activity was defined as producing at least one valid session open and close event during baseline and intervention period.

The final sample consists of 243 participants aged 13–17 years (M = 14.73, Md = 15, SD = 1.44). 
Female participants represent 67.49\% of the sample (n = 164), while male participants represent 32.10\% (n = 78).
Gender data is missing for one participant (0.41\%).
The sample skews slightly young: The largest age group was 13-year-olds (n = 70, 28.81\%), followed by 15-year-olds (n = 50, 20.57\%), 14-year-olds (n = 44, 18.11\%), 16-year-olds (n = 40, 16.56\%), and 17-year-olds (n = 39, 16.05\%). 
Geographical distribution of participants across all five Danish regions corresponds to the general population distribution, with the largest proportions from the capital region (n = 74, 30.45\%), followed by central Denmark (n = 70, 28.81\%), southern Denmark (n = 41, 16.87\%), Zealand (n = 36, 14.81\%), and northern Denmark (n = 22, 9.05\%).


\subsubsection{Adolescents Cohort (G)}
\label{subsubsec:adol-g}
Participants were recruited in the second half of 2024 as new users that just downloaded one sec and hadn’t used the app, yet, and through digital advertisement through social media.

\subsubsection{Alternative Cohort}
\label{subsubsec:adol-alt}
Participants were recruited exclusively through the \textit{one sec} app itself,
via an in-app promotional tile and a short video clip distributed through the
app's official channels and the author's personal social media accounts.
Participation was voluntary and began with completion of the pre-survey. As
\textit{one sec} was available only on iOS at the time, the sample is limited to
iPhone users. The preregistration specified a minimum of 700 participants: 500
new users (100 each for the Control, Commit, Default, Daily and Weekly
conditions) and 200 existing users (100 each for Daily and Weekly). New users were those who enrolled immediately after installing the app and thus had no prior exposure to the intervention; existing users had been using the app for some time and were assigned only to the two experimental conditions in order to avoid frustration effects from withdrawing familiar features. Over 2{,}300
people completed the questionnaire within the first three weeks, but only just over 130 of these were new users, so the preliminary analysis was restricted to existing users who had additionally consented to donating retrospective data. 

\subsubsection{Longitudinal Study (U)}
\label{subsubsec:adol-long-usa}
The fourth study set is not re-analysed but serves as a published reference against which the aggregate usage measures reported in the Result Section are compared. \citet{hamilton_improving_2025} conducted a feasibility and acceptability study of passive mobile sensing with 19 adolescents in the United States (M = 15.84 years, SD = [x]), of whom 68\% were boys and 79\% identified as White. Participants were recruited as part of a larger study on adolescent mental health and were required to own an Android device, since iOS does not permit external applications to collect app usage passively. Data collection took place during the COVID-19 pandemic.

Participants installed the AWARE sensing framework for approximately 31 days \cite{ferreira_aware_2015}. Sensor data were processed using the RAPIDS pipeline \cite{vega_reproducible_2021}. The application-foreground sensor yielded 10,038 hourly observations across 645 unique applications, with a mean data yield of 74.18\% (range 34.8–99.9\%). Participant-hours falling below a 50\% yield threshold were treated as missing rather than as non-use. Acceptability was high: participants reported minimal privacy concerns and largely forgot that the sensing application was installed, although parental security applications and device battery settings interfered with collection for some participants.

\subsection{Data Collection}
\label{subsec:data_collection}
\subsubsection{Adolescents Cohort (D)}
\label{subsubsec:adol-d}
Data were collected using the commercially available \textit{one sec} app especially adapted for the experiment. 
\textit{one sec} is a self-nudging app aiming to reduce smartphone overuse.
Users can define target apps. 
Participants were instructed to link one second with all content-centric social media applications.
As per the Danish Competition and Consumer Authority (2025, \textit{Young Consumers and Social Media}, p.~21), social media applications primarily fall into two categories: chat-based and content-based. This study specifically focused on content-based social media.
For each target app they set up automations triggering when the app is opened or closed. 
When the automations fire, the \textit{one sec} app is called and registers the corresponding opening or closing event with a timestamp for the app.
This automation-driven approach allows fine-grained app usage data collection on iOS with the same, controllable mechanism and does not necessitate the constant background activity of the \textit{one sec} app.
As an intervention, \textit{one sec} can serve the user a self-nudge \cite{gruning_directing_2023, haliburton_longitudinal_2024, reijula_self-nudging_2022} providing design friction and the explicit option to dismiss the app opening.

\subsubsection{Adolescents Cohort (G)}
\label{subsubsec:adol-g}
Data were collected using the same adapted version of the commercially available \textit{one sec} app described above. Participants defined their own target apps and set up automations that call \textit{one sec} whenever one of these apps is opened. In contrast to the Danish cohort, only
opening attempts were recorded: each event carries a timestamp, the name of the target app, and the outcome of the attempt, distinguished as the app being opened, the attempt being aborted by closing the app, or the attempt being dismissed through the intervention screen. No closing events were logged, so session durations are not available for this cohort and all analyses are restricted to frequency and abandonment. Because logging is tied to the user-configured list of target apps, the data cover a
self-selected subset of applications rather than overall device use. Entry and exit surveys were administered around the logging period; as the
present analysis draws exclusively on logged behavioural data, these self-report measures are not considered further.

\subsubsection{Alternative Cohort}
\label{subsubsec:adol-alt}
Enrolment began with a short pre-survey on self-reported usage and satisfaction, repeated after the study period; as the present analysis draws exclusively on logged behavioural data, these self-report measures are not considered further. Behavioural data were collected automatically through an opt-in data donation. For every attempt to open an app configured for \textit{one sec}, the app logged a timestamp, the name of the target app, and whether the attempt was aborted or continued; a corresponding timestamp and app name were logged when the app was closed. Session duration is therefore not a recorded field but the difference between the opening and closing timestamps, and aborted attempts yield no duration. Because logging is tied to the user-configured list of target apps, the data cover a self-selected subset of potentially distracting applications rather than overall device use, and app categories were not recorded and would have to be assigned post hoc. The initial budget and any subsequent changes to it were logged in addition, and participants who had been using the app prior to the study could consent to donating their historical records back to the point of installation along with their in-app configuration at study entry. The study was preregistered on \texttt{aspredicted.org}, and the consent form, study description and questionnaire were reviewed by the ethics committee of the corresponding university in January 2023 without objections.

\subsubsection{Longitudinal Study (U)}
\label{subsubsec:adol-long-usa}
Two features of this cohort constrain the comparison. First, application usage was aggregated into hourly bins, so the reported measures are not directly comparable to session-level statistics. Second, social media applications were identified using researcher-coded categories rather than store taxonomies; the authors report that Google Play Store categories exclude YouTube and yield substantially lower estimates.

\subsection{Study Timeline}
\label{subsec:study_timeline}
\subsubsection{Adolescents Cohort (D)}
\label{subsubsec:adol-d}
Participants were recruited and enrolled throughout September and October of 2024. 
The experiment ran from September to December 2024.
After enrollment, participants filled out the entry survey with basic demographic information.
During a baseline period of two weeks opening and closing interactions with the monitored apps were recorded. 
In a following four week period, participants were randomly assigned to three intervention groups. 
After the intervention period, participants filled out the exit survey.

\subsubsection{Adolescents Cohort (G)}
\label{subsubsec:adol-g}
Participants were recruited in-app and began by completing the entry survey. They were then randomly assigned to one of three intervention
conditions, matching those used during the Danish intervention phase, and their opening attempts were recorded continuously over the following six
weeks. After the intervention period, articipants completed the exit survey. Unlike the Danish study, this cohort has no baseline phase: an
intervention was active from the first recorded event onwards, so no period of unmodified use is observed for these participants.

\subsubsection{Alternative Cohort}
\label{subsubsec:adol-alt}
The study builds on an earlier deployment of \textit{one sec} conducted in
autumn 2021. An analysis of existing wellbeing apps was carried out in November 2022, followed by the design of the intervention and the study conditions.
Study materials were submitted for ethical review in January 2023 and the preregistration was filed before data collection began. Recruitment opened on 24 February 2023, and more than 2.300 people had enrolled within the first three weeks. Each participant contributed six weeks of logged data, with existing users optionally contributing historical records covering a considerably longer and more variable window. A first data extract was taken two weeks after launch; at the time of writing, collection was still ongoing, so any figures drawn from that extract are preliminary. Note that the launch date marks
a structural break in the time series: the intervention screen changed for all participants on 24 February, coinciding with a roughly 40\% drop in average weekly opening attempts between the week before the study and week two, part of which is plausibly a novelty effect.

\subsubsection{Longitudinal Study (U)}
\label{subsubsec:adol-long-usa}
Data collection in the reference study took place over a single continuous period of approximately 31 days per participant, without a baseline or intervention phase, since the study was designed to assess the feasibility and acceptability of passive sensing rather than to evaluate an intervention. Participants completed enrolment procedures and installed the AWARE sensing framework at the outset, after which application usage was recorded passively for the duration of the observation window. Collection occurred during the COVID-19 pandemic, a period in which adolescent schooling and social contact were substantially disrupted; the resulting temporal patterns are therefore not directly comparable to those observed in the present cohorts \cite{hamilton_improving_2025}.

\subsection{Analysis}
\label{subsec:analysis}

\subsubsection{Preprocessing}
Before further analysis all available datasets were preprocessed as follows:

\paragraph{Adolescents Cohort (D)}
Raw event logs captured app openings, closings, and dismissals and their corresponding timestamps. 
To reconstruct sessions, each opening event was matched to the nearest valid closing event for the same app within a short time window, allowing for brief intervening events.
This approach accounts for rapid app switching in iOS and the fact that reopening an app within 60 seconds did not trigger a new intervention. 
The same procedure was applied during baseline and intervention periods to ensure comparability. 
Sessions missing a plausible closing event were excluded. 
The baseline dataset corresponds to all sessions for each participant during the baseline period without interventions. All other sessions are collected in the intervention dataset. The total dataset corresponds to all sessions.

\paragraph{Adolescents Cohort (G)}
Data of participants with less than six weeks data (N = 311) were dropped resulting in 262 participants.

\paragraph{Alternative Cohort (Worldwide)}
With 1,550 consenting users providing 4,724,881 app opening attempt events, some older users dominated the event count. Users with recorded usage durations exceeding the mean + 2 SD were excluded from the analysis, dropping 83 participants with 1,058,449 events and leaving 1,467 users with 3,666,432 events.

\subsubsection{Summary Statistics}
\label{subsubsec:summary_stats}
To situate our datasets within related work, we reproduce summary statistics provided by \citet{hamilton_improving_2025} with the Adolescents Cohort (D), capturing the app opening frequency and duration at hourly and daily temporal resolutions

The \textit{Daily Usage Duration per Participant} was calculated by first summing app-use duration for each participant-day, then averaging these daily values within each participant, and finally averaging across participants, ensuring each participant contributed equally. 
Analogously, \textit{Daily App Openings per Participant} was computed, but using the number of app-opening events per day rather than duration. 
Opening events were attributed only to the day on which the corresponding session started.
U\textit{sage Duration per Participant-Day} corresponds to the mean app-use duration across all participant-days. 
Correspondingly, \textit{App Openings per Participant-Day} was calculated as the mean number of app-opening events across participant-days. 
The \textit{Usage Duration per Participant-Hour} is the mean app-use duration across participant-hours. Sessions spanning hour boundaries were split across the respective hours. 
\textit{App Openings per Participant-Hour} were calculated as the mean number of app-opening events across active participant-hours, with each opening attributed exclusively to the hour in which it occurred.
Sessions spanning day or hour boundaries were split at the relevant boundary so that each portion of the session duration was attributed only to the corresponding day or hour. In contrast, each session was counted only once and assigned to the temporal unit containing its opening timestamp, e.g., a session crossing an hour boundary results in two active participant hours but only accounts for one app opening.

\subsubsection{Micro-Usage}
\label{subsubsec:micro_usage}
We reproduce a micro-usage analysis with the Adolescents Cohort (D) following the methodology of \citet{ferreira_contextual_2014}. We identify a natural break in session durations using the Jenks optimization method \cite{jenks_data_1967, viry_mthhjenkspy_2026}, which partitions observations into two clusters. Sessions with durations below the resulting break point are classified as micro-usage, while those above it are classified as non-micro-usage. We then compare the results from the Danish adolescents dataset with findings reported by others \cite{ferreira_contextual_2014, church_understanding_2015, morrison_large-scale_2018}.

Table \ref{tab:demographics-studies} gives an overview over the study populations and datasets.

\subsubsection{Visualization and Clustering}
\label{subsubsec:visulization_and_clustering}
To answer \textbf{RQ1}, we use heatmaps with 5-minute bins to visualize how app-opening activity is distributed throughout the week, with event counts normalized across the full week. We compare activity patterns across cohorts and illustrate how individual-level patterns can provide additional insights by examining the most average user in the Worldwide users dataset. We further investigate whether users can be grouped according to their weekly activity distributions using k-means clustering.

\subsubsection{Participant Re-identification}
\label{subsubsec:participant_re_identification}
Relating to \textbf{RQ2}, we ask: How identifiable is a person from behavioral traces that might be collected for purposes other than identification? 
Ordinary smartphone behaviors have been shown to contain user-identifying signatures \cite{frank_touchalytics_2012}, which, for example, can be used for authentication \cite{abuhamad_sensor-based_2021} or re-identifying individual users from so-called anonymized \cite{tang_user_2024} and even sparse \cite{de_montjoye_unique_2013}, seemingly harmless \cite{de_montjoye_unique_2015} data, warranting privacy concerns.
\citet{welke_differentiating_2016} and \citet{tu_your_2018}, for example, differentiate users based on their unique sets of used apps. We hypothesize that app usage patterns alone may be sufficient to re-identify users with surprisingly high accuracy from otherwise anonymous traces. To investigate this, we quantify how different statistical, temporal, and app-transition-based representations contribute to individual identifiability.

\paragraph{Re-Identification Experiment Setup}
\label{par:re_identification_experiment_setup}

First, each participant's event trace is sorted chronologically by timestamp and split into two equal halves, yielding a training and a test trace. Second, we calculate the respective feature representations for the training and test traces. Any statistics required for feature calculation, such as the mean of the training feature vectors, are computed exclusively from the training data to prevent data leakage. Third, both training and test feature vectors are centered using the training-set mean. Fourth, the feature vectors are aligned and pairwise compared using cosine similarity, producing a ranking of test traces for each training trace. Finally, we report Top-1 accuracy, Top-5 accuracy, and median rank. Top-1 accuracy denotes the proportion of training traces for which the corresponding user's test trace is ranked first. Top-5 accuracy denotes the proportion for which the corresponding test trace appears among the five highest-ranked candidates. Median rank indicates the median position of the correct test trace across all training traces.

\paragraph{Features}
\label{par:re_identification_features}
The following features were explored in various combinations. Features were concatenated into one vector per participant.

\begin{itemize}
\item \textbf{App.}
The \textit{App} feature corresponds to the app usage distribution of a participant: each value in the vector is the fraction of that participant's recorded events that occurred in a specific app. This feature comes closest to \citet{welke_differentiating_2016} and \citet{tu_your_2018}.

\item \textbf{Transition.}
The \textit{Transition} feature is computed by looking at the sequence of apps for each participant. How often they move from each app to each other app is counted and normalized to proportions.

The \textit{App} and \textit{Transition} features require knowledge of which apps are used by a participant. The following features do not require this information.

\item \textbf{Heatmap.}
The \textit{Heatmap} feature is calculated by counting app opening attempt occurrences in a specified bin size. We determined a 45-minute bin size through grid search between 5-minute and 60-minute bins. The counts are either normalized by week, similar to our heatmap visualizations, or used to calculate the probability of activity in this specific bin.

\item \textbf{Statistics.}
\begin{itemize}
    \item \textbf{Circular statistics.}
    Time of day was represented on a 24-hour circular scale. For each participant, circular mean time and resultant length (concentration) were calculated from the sine and cosine of the corresponding hourly angles, accounting for the cyclical nature of time around midnight.

    \item \textbf{Activity entropy.}
    Shannon entropy \cite{shannon_mathematical_1948} was calculated from the proportion of activity occurring in each hour and normalized by $\log(24)$, yielding a measure from 0 (activity concentrated in one hour) to 1 (activity evenly distributed across all hours).

    \item \textbf{Peak activity.}
    Peak hour was defined as the hour with the largest proportion of recorded activity, with peak share representing the proportion occurring during that hour.
\end{itemize}

\end{itemize}

All analysis and visualization code is available on GitHub\footnote{\fillin{GitHub URL}}.

\subsection{Ethics and Privacy}
\label{subsec:ethics_privacy}
\subsubsection{Adolescents Cohort (D)}
\label{subsubsec:adol-d}

The original study protocol, as well as potential secondary data analysis, was reviewed and approved by the Danish Competition and Consumer Authority's internal governance structure (legal department and agency management). 
The approval of a regional research ethics committee was not required under Danish regulations, because the study does not constitute biomedical research or involve human biological material and was assessed as posing no more than minimal risk.
All participants provided informed consent. 
For participants aged 15 or younger, a parent or legal guardian provided written informed consent prior to participation in accordance with Danish law. 
Participants were informed that the study examined their use of digital apps over eight weeks. 
Participation was voluntary. 
No deception was used, and withdrawal from the study was possible at any time without consequences. 
The compensation was disclosed during recruitment and consent. The gift card was provided upon completion of the study.  Due to a technical error, app usage data were recorded only during the first six, rather than the planned eight, weeks of the study. 
This did not affect the compensation or rights of participants. Data processing complied with Danish data protection legislation and the EU General Data Protection Regulation (GDPR). The commercially available \textit{one sec} app is self-financed and does not share user data with third parties for profit \cite{haliburton_longitudinal_2024}.

\subsubsection{Adolescents Cohort (G)}
\label{subsubsec:adol-g}
The study was assessed internally at the institution of one of the authors (\fillin{institution}), which determined that it did not require a formal ethics vote, on the grounds that the data collected are neither sensitive nor suitable for identifying individuals.

\subsubsection{Alternative Cohort}
\label{subsubsec:adol-alt}
The study was preregistered on \texttt{aspredicted.org}. The consent form, study description and questionnaire were submitted to the ethics committee of the corresponding university in January 2023, which raised no objections. Participation was voluntary throughout, and all behavioural data were collected through an explicit opt-in data donation; users who had installed the app before the study could separately consent to donating their historical records.

\subsubsection{Longitudinal Study (U)}
\label{subsubsec:adol-long-usa}
All study procedures in the reference study were approved by the institutional review board. Participants aged 18 provided informed consent; those aged 13–17 provided assent alongside parental consent, obtained in a video call with study staff. Participants did not consent to data sharing, so the analysis reported here uses only the published aggregate statistics.

\section{Results}
\label{sec:results}
\subsection{Summary Statistics}

\begin{table}[htbp]
    \centering
    \small
    \setlength{\tabcolsep}{3pt}
    \caption{Summary statistics for daily and hourly activity measures (duration in minutes and opening frequency) for the Adolescents Cohort (D) by study period.}
    \label{tab:summary_statistics_danish_adolescents}
\resizebox{\textwidth}{!}{%
\begin{tabular}{lrrrrrrrrrrrrrrr}
\toprule
 & \multicolumn{5}{c}{Baseline} & \multicolumn{5}{c}{Intervention} & \multicolumn{5}{c}{Total} \\
\cmidrule(lr){2-6}
\cmidrule(lr){7-11}
\cmidrule(lr){12-16}
Metric & M & SD & Min & Max & N & M & SD & Min & Max & N & M & SD & Min & Max & N \\
\midrule
Daily Usage Duration per Participant & 179.50 & 121.10 & 1.10 & 622.32 & 243 & 131.97 & 97.57 & 0.20 & 565.20 & 241 & 153.19 & 103.15 & 1.37 & 597.52 & 243 \\
Daily App Openings per Participant & 44.21 & 35.41 & 1.00 & 191.13 & 243 & 28.78 & 25.54 & 1.20 & 157.97 & 241 & 35.28 & 28.49 & 1.00 & 168.64 & 243 \\
Usage Duration per Participant-Day & 183.97 & 150.69 & 0.01 & 822.81 & 3,500 & 137.04 & 120.63 & 0.02 & 890.31 & 6,520 & 156.61 & 134.37 & 0.01 & 890.31 & 9,817 \\
App Openings per Participant-Day & 45.56 & 41.42 & 1.00 & 333.00 & 3,500 & 29.81 & 29.00 & 1.00 & 287.00 & 6,520 & 36.04 & 34.90 & 1.00 & 333.00 & 9,817 \\
Usage Duration per Participant-Hour & 16.94 & 17.29 & 0.00 & 60.00 & 38,008 & 13.73 & 15.10 & 0.00 & 60.00 & 65,072 & 14.93 & 16.03 & 0.00 & 60.00 & 102,992 \\
App Openings per Participant-Hour & 4.20 & 4.21 & 0.00 & 39.00 & 38,008 & 2.99 & 3.01 & 0.00 & 45.00 & 65,072 & 3.44 & 3.55 & 0.00 & 45.00 & 102,992 \\
\bottomrule
\end{tabular}
}
\end{table}

\begin{table}[htbp]
    \centering
    \small
    \setlength{\tabcolsep}{3pt}
    \caption{Instagram only: Summary statistics for daily and hourly activity measures (duration and checking) for Danish Adolescents by study period.}
    \label{tab:summary_statistics_danish_adolescents_instagram}
\resizebox{\textwidth}{!}{%
\begin{tabular}{lrrrrrrrrrrrrrrr}
\toprule
 & \multicolumn{5}{c}{Baseline} & \multicolumn{5}{c}{Intervention} & \multicolumn{5}{c}{Total} \\
\cmidrule(lr){2-6}
\cmidrule(lr){7-11}
\cmidrule(lr){12-16}
Metric & M & SD & Min & Max & N & M & SD & Min & Max & N & M & SD & Min & Max & N \\
\midrule
Daily Usage Duration per Participant & 36.81 & 37.55 & 0.03 & 204.41 & 199 & 27.95 & 28.96 & 0.02 & 162.08 & 198 & 32.07 & 31.63 & 0.22 & 176.20 & 199 \\
Daily App Openings per Participant & 11.34 & 9.77 & 1.00 & 72.25 & 199 & 7.08 & 6.52 & 1.00 & 49.21 & 198 & 8.85 & 7.58 & 1.00 & 58.93 & 199 \\
Usage Duration per Participant-Day & 39.81 & 52.56 & 0.01 & 388.32 & 2,580 & 31.78 & 43.80 & 0.00 & 378.24 & 4,645 & 35.20 & 47.62 & 0.00 & 388.32 & 7,111 \\
App Openings per Participant-Day & 12.50 & 12.44 & 1.00 & 115.00 & 2,580 & 8.04 & 8.10 & 1.00 & 77.00 & 4,644 & 9.79 & 10.18 & 1.00 & 115.00 & 7,110 \\
Usage Duration per Participant-Hour & 6.36 & 10.35 & 0.00 & 60.00 & 16,158 & 6.16 & 9.69 & 0.00 & 60.00 & 23,944 & 6.24 & 9.96 & 0.00 & 60.00 & 40,084 \\
App Openings per Participant-Hour & 2.00 & 1.70 & 0.00 & 27.00 & 16,158 & 1.56 & 1.14 & 0.00 & 14.00 & 23,944 & 1.74 & 1.41 & 0.00 & 27.00 & 40,084 \\
\bottomrule
\end{tabular}
}
\end{table}

Table \ref{tab:summary_statistics_danish_adolescents} summarizes app usage across the baseline and intervention periods for the adolescence (D) cohort. On average participants spend 2 hours and 59 minutes using social media with on average 3.65 apps used from the tracked apps. Snapchat is the app tracked by the most participants. Overall, usage was lower during the intervention period across all measures. At the participant level, mean daily usage duration decreased from 179.50 min (SD = 121.10) at baseline to 131.97 min (SD = 97.57) during the intervention, while mean daily app openings decreased from 44.21 (SD = 35.41) to 28.78 (SD = 25.54). Similar reductions were observed at the participant-day level, with mean usage duration decreasing from 183.97 to 137.04 min and mean app openings from 45.56 to 29.81.

Hourly measures showed the same pattern. Mean usage duration per participant-hour decreased from 16.94 min (SD = 17.29) at baseline to 13.73 min (SD = 15.10) during the intervention, while app openings decreased from 4.20 (SD = 4.21) to 2.99 (SD = 3.01). The intervention period also exhibited substantial between-observation variability, with usage duration ranging from near zero to 890.31 min per participant-day and from 0 to 60 min per participant-hour. Across the full observation period, the corresponding means were 156.61 min per participant-day and 14.93 min per participant-hour, with 36.04 and 3.44 app openings, respectively. These descriptive statistics indicate a consistent reduction in both the duration and frequency of app engagement during the intervention period.

In Longitudinal Study (U) \cite{hamilton_improving_2025} a mean daily usage duration per participant of 94.4 min (SD = 62.47), a mean usage duration per participant-day of 98.05 min (SD = 79.34), and a mean usage duration per participant-hour of 5.51 min (SD = 9.11) are reported. Baseline duration values are nearly double. In comparison, Longitudinal Study (U) reports significantly higher average app openings with mean daily app openings per participant of 159.97 (SD = 125.30), mean app openings per participant day of 164.15 (SD = 146.30) and mean app openings per participant-hour of 9.22 (14.21).

As an example for analyzing app data by specific app, we include summary statistics for the Adolescents Cohort (D) and the app Instagram in Table~\ref{tab:summary_statistics_danish_adolescents_instagram}.

\begin{table}[htbp]
\centering
\small
\setlength{\tabcolsep}{4pt}
\caption{Summary statistics for daily and hourly activity app opening attempts for the Alternative Cohort (Worldwide) and the Adolescents Cohort (G). Session durations cannot be reported due to the lack of closing events in the data.}
\label{tab:summary_statistics_german_adolescents_worldwide}

\begin{tabular}{lrrrrr|rrrrr}
    \toprule
    & \multicolumn{5}{c|}{Alternative Cohort (Worldwide)} 
    & \multicolumn{5}{c}{Adolescents Cohort (G)} \\
    \cmidrule(lr){2-6}
    \cmidrule(lr){7-11}
    Metric & M & SD & Min & Max & N 
    & M & SD & Min & Max & N \\
    \midrule
    
    Daily App Openings per Participant
    & 25.31 & 21.31 & 1.40 & 144.69 & 1,467
    & 18.38 & 14.82 & 2.00 & 96.53 & 262 \\
    
    App Openings per Participant-Day
    & 27.36 & 28.80 & 1.00 & 1043.00 & 133,881
    & 19.08 & 19.26 & 1.00 & 249.00 & 10,528 \\
    
    App Openings per Participant-Hour
    & 3.58 & 3.81 & 1.00 & 509.00 & 1,023,109
    & 2.86 & 2.80 & 1.00 & 86.00 & 70,254 \\
    
    \bottomrule
\end{tabular}
\end{table}

Table~\ref{tab:summary_statistics_german_adolescents_worldwide} contains summary statistics for app opening attempts in the Alternative (Worldwide) and Adolescents (G) cohort. Counts are overall lower than in the Adolescents (D) and Longitudinal Study (U), since per participant in cohort (D) only 1.79 and per participant in the worldwide cohort only 3.52 apps are tracked on average. The most tracked app in both datasets is Instagram, which may result in fewer app opening attempts than a more frequented app like Snapchat as explored in the next section.

\subsection{Micro Usage}
\begin{figure}[h]
  \centering
  \includegraphics[width=0.48\columnwidth]{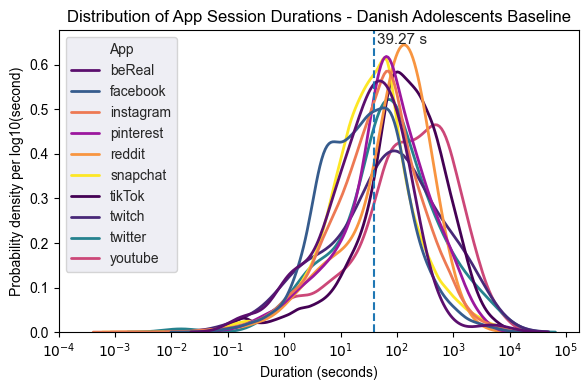}
  \caption{Kernel Density Estimate (KDE) plot showing the probabilities of app session duration per social media app in the baseline usage data of Danish adolescents. The natural break was calculated using Jenks optimization method~\cite{jenks_data_1967, viry_mthhjenkspy_2026} following previous studies~\cite{ferreira_contextual_2014, church_understanding_2015, morrison_large-scale_2018} to discriminate between micro-usage below and non-micro-usage above the threshold. Micro-usage was detected for 66,083 sessions and non-micro-usage for 93,366 sessions.}
  \Description{Kernel density curves of app session duration for ten social media apps, plotted on a logarithmic duration axis. The horizontal axis runs from 0.0001 to 100,000 seconds and the vertical axis gives probability density per log-10 second, from 0 to about 0.65. Each app has its own coloured curve, identified in the legend. A dashed vertical line marks the 39.27-second threshold. All curves are broadly unimodal and overlap substantially. Most peak between roughly 60 and 150 seconds, to the right of the threshold. Facebook is the clearest exception, peaking near 10 seconds, well to the left of the threshold. YouTube peaks furthest right, at several hundred seconds. Density is negligible below 0.1 seconds and above 10,000 seconds for every app.}\label{fig:danish_adolescents_baseline_app_usage_session_duration_distribution}
\end{figure}

\begin{figure}[h]
  \centering
  \includegraphics[width=0.70\columnwidth]{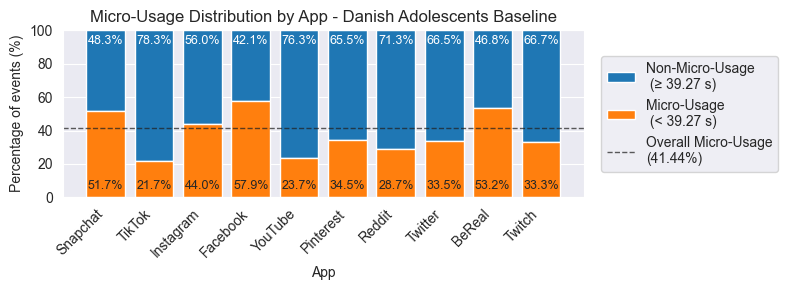}
  \caption{Micro-usage distribution by social media app for the Danish adolescent baseline data. The natural break was calculated using Jenks optimization method~\cite{jenks_data_1967, viry_mthhjenkspy_2026} following previous studies~\cite{ferreira_contextual_2014, church_understanding_2015, morrison_large-scale_2018} to discriminate between micro-usage below and non-micro-usage above the threshold. Overall 41.44\% of app usage is micro-usage. Apps are sorted by session count from left to right. Note, the session count is dominated by three apps (Snapchat (N = 74,195), TikTok (N = 41,831), and Instagram (N = 32,247)). Thus, despite having more non-micro-usage, Twitch (N = 39) only contributes to fewer overall app usage duration. }
  \Description{Stacked bar chart showing the split between micro-usage and longer sessions for ten social media apps. Each bar totals 100 percent of that app's events, with the micro-usage share at the bottom and the remainder above; a dashed horizontal line marks the overall micro-usage rate of 41.44 percent. Apps are ordered left to right by session count. Micro-usage shares are: Snapchat 51.7, TikTok 21.7, Instagram 44.0, Facebook 57.9, YouTube 23.7, Pinterest 34.5, Reddit 28.7, Twitter 33.5, BeReal 53.2, and Twitch 33.3 percent. Facebook, Snapchat, and BeReal are the only apps above 50 percent; TikTok and YouTube are the lowest, at roughly a fifth to a quarter..}
  \label{fig:danish_adolescents_baseline_app_usage_session_duration_distribution_micro_usage_by_app}
\end{figure}

The natural break calculated using Jenks optimization method was found at 39.27s for the Adolescents Cohort (D), resulting in a percentage of micro-usage of 41.44\%.
The break point is higher than the break points found for three datasets by \citet{church_understanding_2015}, who report values of 16.6s, 22.5s, and 21.5s, corresponding to 53\%, 56\%, and 55\% of micro-usage. Reporting on iOS users, \citet{morrison_large-scale_2018} identfies the break at 21.4s with 43.6\% of usage classed as micro-usage. \citet{ferreira_contextual_2014} identify the break at about 15s counting 41.5\% as micro-usage. 
A probability distribution function estimation for session duration can be found in Figure~\ref{fig:danish_adolescents_baseline_app_usage_session_duration_distribution} for each app in the dataset. Similar, to the other authors \cite{ferreira_contextual_2014, church_understanding_2015, morrison_large-scale_2018} we find micro-usage not to occur equally within the apps. Figure~\ref{fig:danish_adolescents_baseline_app_usage_session_duration_distribution_micro_usage_by_app} shows the micro-usage distribution by app. YouTube and TikTok have the fewest micro-usage, whereas Snapchat, Facebook and BeReal show more than average micro-usage. This is potentially indicative of checking and sending short messages versus engaging longer with video content.

Exploring the break identification using k-means (k = 2) similar to \citet{church_understanding_2015} and \citet{morrison_large-scale_2018} we identify slightly different break points in our data compared to Jenks optimization method, with a break point at 40.01s corresponding to 41.84\% micro-usage. Both studies do not report how the break is exactly calculated: Is it the mean between the highest micro-usage and lowest non-micro-usage observation or the mean between the centroids of the micro-usage and non-micro-usage clusters? On our data both methods yielded the same break point, only differing in the fourth decimal place.

\subsection{Visualization and Clustering}
\label{subsec:viz_clustering}

\subsubsection{Usage Pattern Visualization}
\label{subsubsec:results_usage_pattern_visualization}
\begin{figure}[h]
  \centering
 \includegraphics[width=\linewidth]{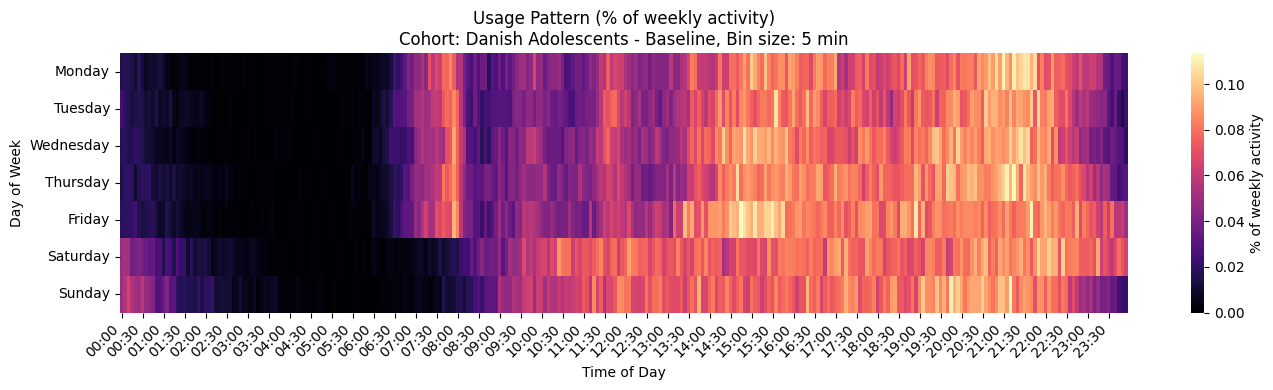}
  \caption{Social media usage pattern as the percentage of weekly activity for Danish adolescents during the baseline period calculated using only opening events. Bin size is 5 minutes and the timezone is Copenhagen.}
  \Description{Heatmap of social media activity by day of week and time of
day for Danish adolescents during baseline. Seven rows, Monday at the top
to Sunday at the bottom, are plotted against time of day from 00:00 to
23:55 in five-minute columns. Cell brightness encodes the share of weekly
activity, from dark at zero to bright at about 0.11 percent. A wide dark
band covers roughly 23:30 to 06:30 on weekdays. On each weekday a narrow
bright column appears at about 08:00, followed by alternating darker and
brighter vertical stripes through the late morning. Brightness increases
from early afternoon and peaks between about 19:00 and 22:00, with the
single brightest region on Monday evening and a further bright block on
Friday between about 14:00 and 16:00. Saturday and Sunday show no 08:00
column, a later and more gradual morning onset, activity spread more
evenly across the day, and activity continuing past midnight.}
  \label{fig:danish_adolescents_baseline_weekly_pattern}
\end{figure}

We first explore the heatmap visualization of social media usage pattern in adolescents cohort (D). 
The baseline pattern on the cohort level in Figure~\ref{fig:danish_adolescents_baseline_weekly_pattern} shows a clear activity difference between day and night time, indicating sleep times. During the week usage ramps up around 6:30 am, most likely corresponding to wake times. A pek can be observed just before school starts at 8 am.
Recess times are indicated by lower peaks and 90-minute modules, corresponding to typical lesson lengths, can be observed as activity dips in the morning. Increased usage activity happens around noon and in the evening. The peak at 3 pm on Fridays corresponds with the popular "Friday bar" ("Fredagsbar") or "Friday café" (Fredagscafé) events at danish learning institutions, where students socialize.

Patterns differ between weekday and weekend: Adolescents stay active longer on Fridays and Saturdays. They return to their evening schedule on Sunday to possibly prepare for waking up early on Mondays. Social Media use is also more equally distributed over the whole day on the weekend compared to more clustered use during weekdays.

The highly structured activity patterns are most likely due to participants in cohort (D) being all school aged and Danish schools times being similar throughout the country, i.e. the cohort is rather homogeneous.

\begin{figure}[h]
  \centering
 \includegraphics[width=\linewidth]{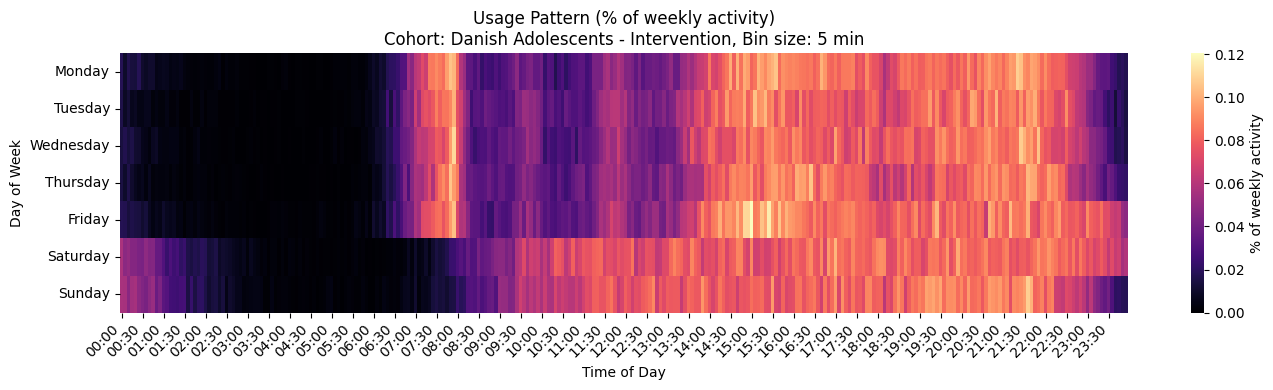}
  \caption{Social media usage pattern as the percentage of weekly activity for Danish adolescents during the intervention period calculated using only opening events. Bin size is 5 minutes. }
  \Description{Heatmap of social media activity by day of week and time of
day for Danish adolescents during the intervention period. Axes and colour
encoding match the previous figure, with the scale reaching about 0.12
percent. The overall structure is retained: a dark overnight band, a
bright column near 08:00 on weekdays, and afternoon and evening peaks.
Compared with the baseline figure, transitions between adjacent
five-minute columns are smoother and less speckled, and the brightest
cells fall on Friday in the mid-afternoon and on weekday evenings.}
  \label{fig:danish_adolescents_intervention_weekly_pattern}
\end{figure}

\begin{figure}[h]
  \centering
 \includegraphics[width=\linewidth]{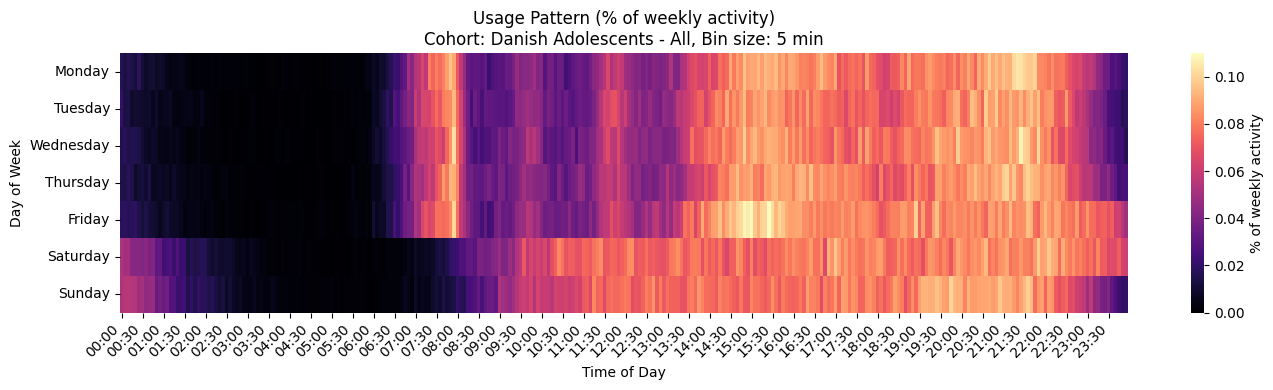}
  \caption{Social media usage pattern as the percentage of weekly activity for Danish adolescents during the whole experiment (baseline and intervention period) calculated using all opening events. Bin size is 5 minutes. Note, how more events recorded during the intervention period leads to smoother color gradients.}
  \Description{Heatmap of social media activity by day of week and time of
day for Danish adolescents across the whole study. Axes and colour
encoding match the two previous figures, with the scale reaching about
0.10 percent. The combined data show the same overnight trough, weekday
08:00 column, and afternoon-to-evening peak, with smoother gradients than
either period alone and the weekend rows again lacking the morning
column.}
  \label{fig:danish_adolescents_all_weekly_pattern}
\end{figure}

The cohort level app opening  pattern stays intact even under intervention as shown in Figure~\ref{fig:danish_adolescents_intervention_weekly_pattern}. Peaks, dips, and overall weekly rhythms do not change. This can be additionally observed in Figure~\ref{fig:danish_adolescents_all_weekly_pattern}, where all intervention and baseline openings are visualized, resulting in a visually nearly identical pattern.
Note, how more events visualized lead to visually smoother gradients.

\begin{figure}[h]
  \centering
 \includegraphics[width=\linewidth]{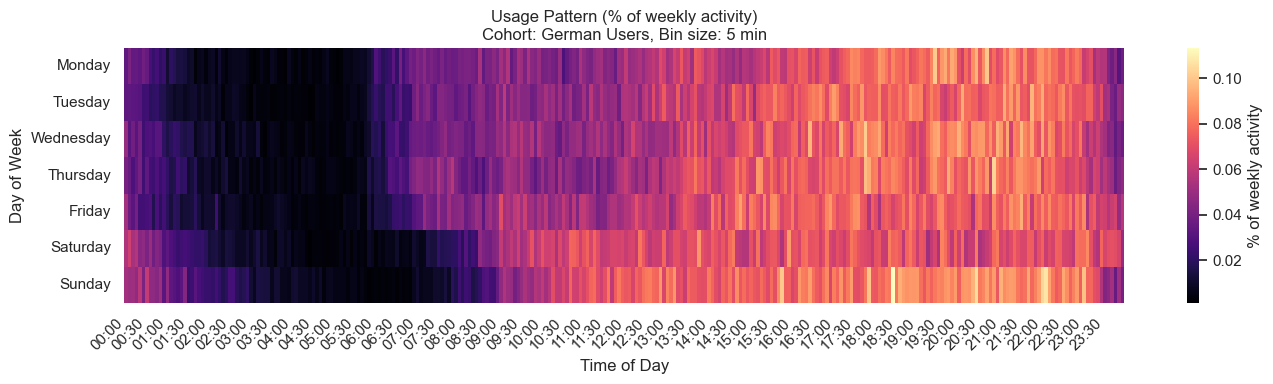}
  \caption{App opening attempt pattern as the percentage of weekly activity for the adolescents cohort (G). Bin size is 5 minutes.}
  \Description{Heatmap of social media activity by day of week and time of
day for German users. Axes match the earlier heatmaps and the colour scale
runs from about 0.02 to 0.10 percent of weekly activity. A dark band
covers roughly 00:00 to 06:00, narrower and less uniformly dark than in the Danish figures. Brightness rises gradually through the morning without any sharp column, remains moderate across the middle of the day, and reaches its maximum between about 18:00 and 21:00. Saturday and Sunday are brighter than weekdays in the late-night and early-morning hours, and the
Sunday evening peak is the brightest region in the figure. Within-day
variation is visibly weaker than in the Danish adolescent figures.}
  \label{fig:german_users_weekly_pattern}
\end{figure}

The app opening attempt pattern of the German adolescents cohort in Figure~\ref{fig:german_users_weekly_pattern} again exhibits a distinct night and day difference. Activity starts at a similar time to the Danish adolescents. Similarly, less usage occurs during the first half of the day and more usage in the afternoon and evening hours. In comparison to the Danish Adolescents, less structure within each day is observable and usage fades out longer after midnight, indicating going to bed later. This may be due to the cohort on average being older than the Danish cohort, i.e., fewer participants follow a strict school schedule. Additionally, German school times are more hetergenous than in Denmark.

The usage patterns still differ between weekend and non-weekend days. The general day-night usage pattern shape stays is very similar, with weekdays usage starting around 7 am and weekend usage around 8:30 am on Saturdays and 9:30 am on Sundays. The start of the day is, in comparison to the Danish adolescents, not marked by intense activity around 8 am.

Figure~\ref{fig:german_users_weekly_pattern} shows the app opening attempt pattern of the alternative cohort of worldwide users. 
The night period is less distinct in comparison to the Danish and German adolescents. In comparison to the Danish Adolescents, less structure within each day is observable. Activity is low in the morning and ramps up in the afternoon to peak in the evening. This is in line with findings of prior studies \cite{morrison_large-scale_2018}
The usage patterns still differ between weekend and non-weekend days. Usage concentrates around the afternoon and evening hours and ends later on Saturday and Friday compared to other days. 
The gradients in the figure are smoother due to more events being visualized.
Unfortunately, timezone-specific analysis cannot be performed, due to the absence of timezone information in the raw data and only UTC timestamps being available. This could be one reason for a less distinctly visible day-night pattern.

\begin{figure}[h]
  \centering
 \includegraphics[width=\linewidth]{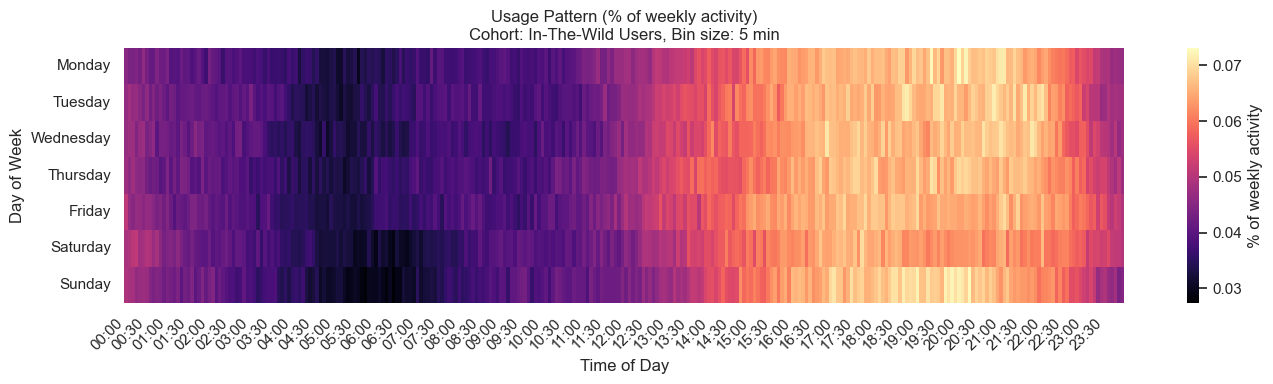}
  \caption{App opening attempt pattern as the percentage of weekly activity for the Alternative Cohort (Worldwide). Bin size is 5 minutes using UTC timestamps.}
  \Description{Heatmap of social media activity by day of week and time of
day for in-the-wild users, using coordinated universal time. Axes match
the earlier heatmaps, but the colour scale spans only about 0.03 to 0.07
percent, so contrast across the figure is much lower. The darkest region
falls between roughly 04:00 and 07:00 and no part of the figure is fully
dark. Brightness rises through the early afternoon and stays high from
about 15:00 to 23:00 on all seven days. Saturday and Sunday are somewhat
brighter than weekdays in the early morning. No sharp within-day
transitions are visible.}
\label{fig:in_the_wild_users_intervention_weekly_pattern}
\end{figure}

\begin{figure}[h]
  \centering
 \includegraphics[width=\linewidth]{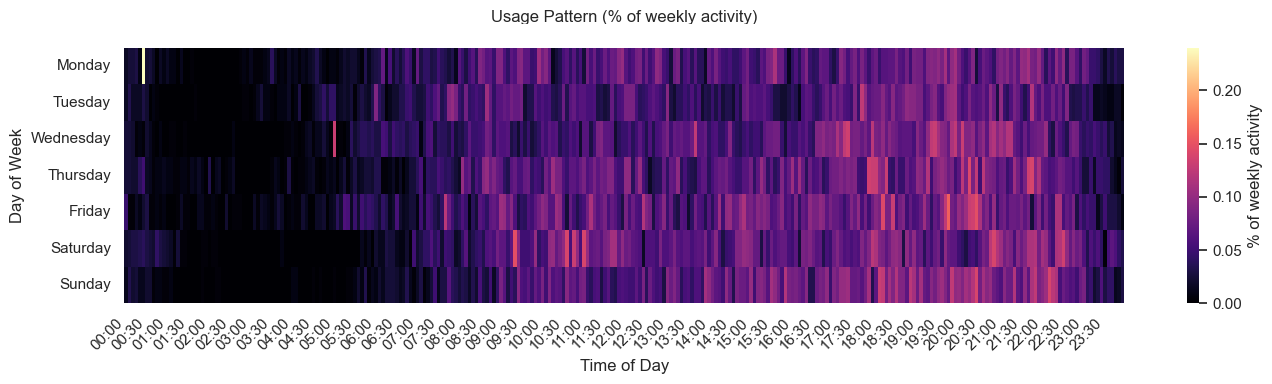}

  \caption{App opening pattern as the percentage of weekly activity of the most average participant in the alternative cohort (worldwide), calculated displaying their 23,753 opening attempt events. Bin size is 5 minutes and the time is in UTC.}
  \label{fig:most_average_user_worldwide_users}
  \Description{Heatmap of one participant's social media activity by day of
week and time of day. Axes match the cohort heatmaps and the colour scale
runs from zero to about 0.22 percent of that participant's weekly
activity. Because the data come from one person rather than a cohort, most
cells are dark or empty and activity appears as scattered isolated
columns rather than continuous bands. A single very bright column appears
on Monday at about 00:30, and further isolated bright cells occur on
Wednesday near 05:20 and on Friday in the evening. Moderate activity
clusters in the late afternoon and evening on most days, and a dark region
covers roughly 01:00 to 07:00.}
  \end{figure}

App opening patterns can not only be visualized on a cohort-level, but also on an individual basis. To demonstrate this, we identified the most average participant in the alternative cohort of worldwide participants as the participant closest to the mean of all usage heatmaps in the dataset using the Euclidean distance. the participant's app opening pattern is shown in Figure~\ref{fig:most_average_user_worldwide_users}. 
In comparison to cohort-level visualization, the heatmap is less smooth, most likely caused by fewer events (N = 23,753) being visualized. Sleeping periods at night and a gradual onset of activity in the morning can be identified. A difference between weekdays and weekends is again visible. The high percentage of activity in the early hours on Monday may indicate erroneous logging, noise, or a distinct notification pattern causing a regular usage spike. This raises the question: should data cleaning remove this seemingly unnatural observation clustering or is it indicative for a specific behavior of this user?

\subsubsection{Clustering}
\label{subsubsec:results_clustering}
\begin{figure}[h]
  \centering
  \includegraphics[width=0.49\columnwidth]{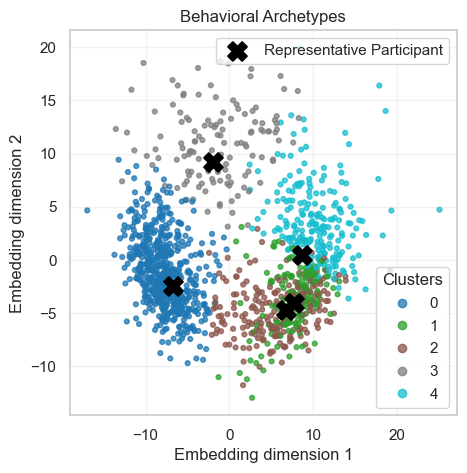}
  \caption{K-means ($k = 5$) clustering of the participants in the alternative cohort (worldwide) in 15-minute bins normalized by week and visualized in 2-dimensional embedding space using PCA.}
  \Description{Scatter plot of participants in a two-dimensional embedding, coloured by k-means cluster, with five clusters. Both axes are unitless embedding dimensions. The horizontal axis runs from about minus 17 to 22 and the vertical axis from about minus 12 to 20. Each small point is one participant and colour denotes cluster membership, numbered 0 to 4 in the legend. Five black crosses mark the representative participant of each cluster. Cluster 0 forms a dense, clearly separated group on the left, centred near minus 7 on the horizontal axis. Cluster 3 sits above and right of it, centred near minus 2 and 9, and is more diffuse. Clusters 1, 2 and 4 occupy the right-hand side and overlap substantially: cluster 4 spreads across the upper right, cluster 2 forms a denser band below it, and cluster 1 is the smallest and is scattered within and around cluster 2. The representative participants for clusters 1 and 2 lie close together near 8 on the horizontal axis, reflecting that overlap.}
  \label{fig:worldwide_users_kmeans_clustering}
\end{figure}

\begin{figure}[h]
  \centering
  \includegraphics[width=\columnwidth]{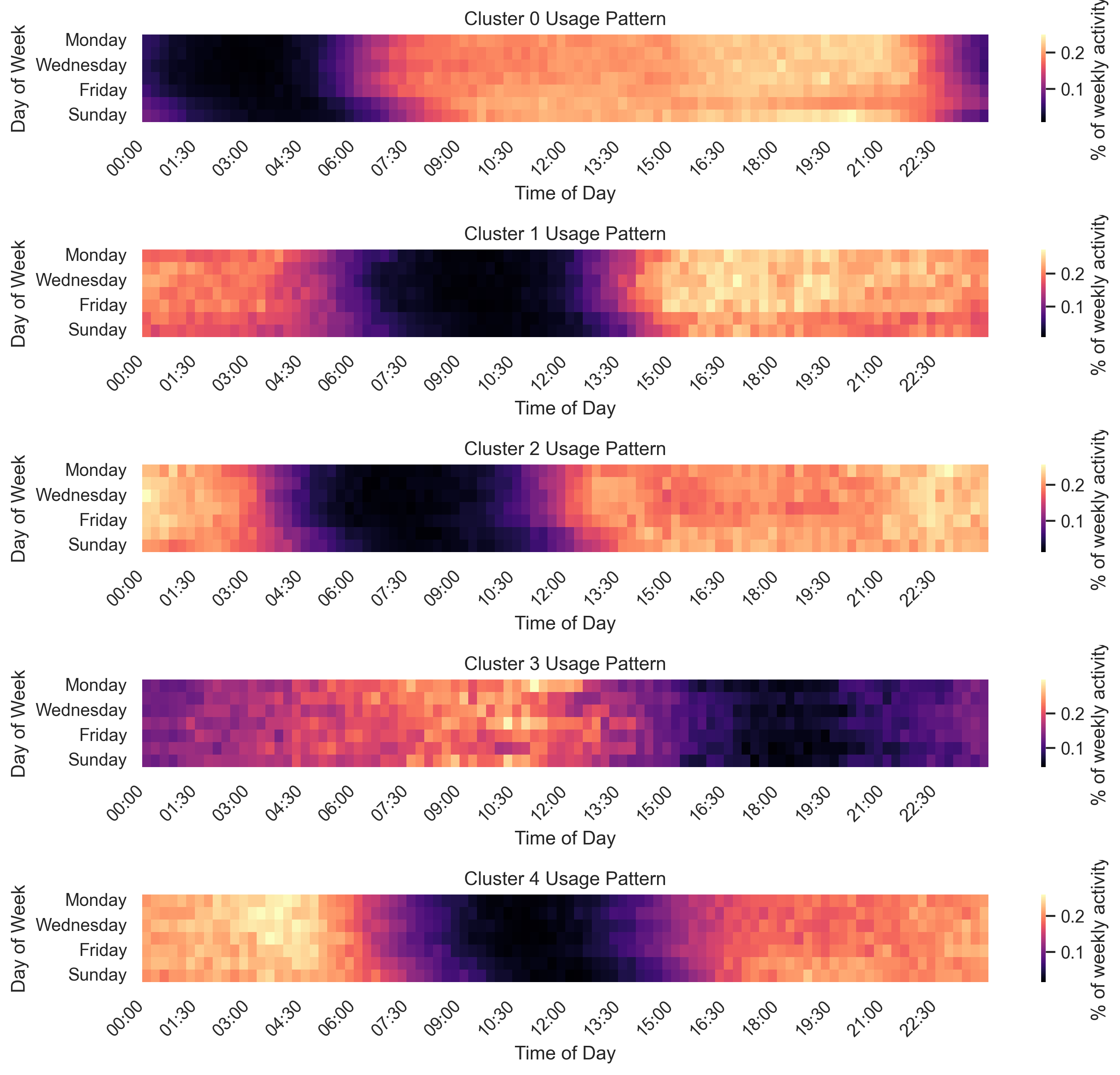}
  \caption{Usage heatmaps visualizing the clusters resulting from the K-means ($k = 5$) clustering of users in the worldwide dataset in 15-minute bins normalized by week. All time is in UTC.}
  \Description{Five stacked heatmaps, one per cluster, showing when each group of participants is active across the week. Each panel plots day of week vertically, labeled at Monday, Wednesday, Friday and Sunday, against time of day from 00:00 to 23:30 horizontally. Cell brightness encodes the share of that cluster's weekly activity on a common scale from zero to about 0.25 percent, dark for low and bright for high. Within every panel the rows resemble one another, so the clusters differ mainly in time of day rather than in day of week. The five panels separate as follows. Cluster 0 is dark from midnight until about 06:00, brightens through the morning, stays uniformly bright from roughly 09:00 to 22:30, and drops sharply after 23:00. Cluster 1 is moderately bright overnight until about 04:30, dark from 06:00 to 13:30, and brightest between 14:30 and 22:00, remaining active past midnight. Cluster 2 resembles cluster 1 but with its dark period shifted earlier, from about 04:30 to 12:00, and its brightest cells in the late evening after 22:00. Cluster 3 is the inverse of the others: its brightest region falls between 07:30 and 13:30, peaking near 11:00, and it is dark from 15:00 through the rest of the day. Cluster 4 peaks overnight between about 02:00 and 05:00, is darkest from 09:00 to 13:30, and recovers to moderate brightness from late afternoon onwards. }
  \label{fig:all_cluster_usage_heatmaps}
\end{figure}

The k-means (k= 5) clustering of the participants in the alternative cohort (worldwide) by their app opening pattern normalized by week in 15 minute bins results can be seen in Figure~\ref{fig:worldwide_users_kmeans_clustering}. Five distinct clusters were identified among the 1,325 participants having a minimum of 200 events. Cluster 0 was the largest group, comprising 591 observations (44.6\%). Clusters 2 and 4 included 223 (16.8\%) and 224 (16.9\%) observations, respectively, while Cluster 1 contained 156 observations (11.8\%). Cluster 3 was the smallest cluster, with 131 observations (9.9\%).
We tuned $k$ to result in visually meaningful activity heatmaps per cluster displayed in Figure~\ref{fig:all_cluster_usage_heatmaps}. The patterns mainly differ between clusters in their period of inactivity. Since the cohort data does not include timezone information, the clusters most likely correspond to the main timezones in the dataset. Note, how the shape of the hotspots exhibits the characteristic shape difference between weekdays and weekend similar to the heatmaps of the other cohorts explored above, just shifted by a few hours.

\subsection{Participant Re-Identification}
\label{subsec:results_re-identification}

\subsubsection{Participant Re-identification}
\label{subsubsec:participant_re_identification}

The results in Table~\ref{tab:reidentification_results} show that participants can be re-identified substantially better than chance using behavioral traces alone. Across all cohorts, the strongest performance was obtained with the combination of \textit{App}, \textit{Heatmap}, and \textit{Transition} features, whereas adding the circular statistics did not improve identification. For the Adolescents Cohort (D), this configuration achieved a Top-1 accuracy of 22.22\% and a Top-5 accuracy of 39.51\%, with a median rank of 11. In Cohort (G), it achieved 21.76\% Top-1 and 41.98\% Top-5 accuracy, with a median rank of 8. The same configuration performed somewhat lower in the larger Alternative Cohort, reaching 15.27\% Top-1 and 29.93\% Top-5 accuracy, with a median rank of 30.

Temporal activity patterns alone were also informative. The \textit{Heatmap} feature using activity probabilities achieved Top-1 accuracies of 18.52\%, 13.74\%, and 8.32\% for Cohorts (D), (G), and the Alternative Cohort, respectively. In contrast, the week-normalized heatmap generally performed worse, particularly for the adolescent cohorts. Adding the statistical features to the activity-probability heatmap did not improve performance and generally reduced both Top-1 and Top-5 accuracy.

All feature configurations substantially outperformed the random baseline. For example, the best Top-1 accuracies were approximately 51, 57, and 218 times higher than random chance for Cohorts (D), (G), and the Alternative Cohort, respectively. Overall, these results indicate that app usage and app-transition patterns contain considerable individual-specific information, while temporal activity patterns alone already provide a substantial degree of re-identifiability.

\begin{table*}[t]
\centering
\caption{Re-identification performance across user cohorts and feature
configurations. For each cohort, Top-1 accuracy, Top-5 accuracy, and median
rank are reported for each evaluated combination of app usage, temporal
heatmap, statistical, and transition features. Heatmap representations use a
45-minute bin size throughout the
experiment. The random baseline is reported separately for each cohort as a
reference for identification based on random re-identification of a user. Higher Top-1 and Top-5 accuracy as well as higher median rank indicates better performance. Best results per cohort are marked in bold.}
\label{tab:reidentification_results}

\small
\setlength{\tabcolsep}{4pt}

\begin{tabularx}{\textwidth}{
    >{\raggedright\arraybackslash}p{0.13\textwidth}
    >{\raggedright\arraybackslash}X
    >{\raggedright\arraybackslash}p{0.15\textwidth}
    >{\raggedright\arraybackslash}p{0.16\textwidth}
    >{\centering\arraybackslash}p{0.09\textwidth}
    >{\centering\arraybackslash}p{0.09\textwidth}
    >{\centering\arraybackslash}p{0.10\textwidth}
}
\toprule
\textbf{Cohort}
& \textbf{Feature configuration}
& \textbf{Heatmap metric}
& \textbf{Statistics}
& \textbf{Top-1} $\uparrow$
& \textbf{Top-5} $\uparrow$
& \textbf{Median Rank} $\uparrow$ \\
\midrule

\multirow{6}{*}{\parbox[t]{0.13\textwidth}{Adolescents Cohort (D)\\($n=243$)}}
& App + Heatmap + Statistics + Transition
& Activity probability
& Circular
& 20.99\%
& \textbf{40.74\%}
& \textbf{8.0} \\

& App + Heatmap + Transition
& Activity probability
& --
& \textbf{22.22\%}
& 39.51\%
& 11.0 \\

& Heatmap
& Activity probability
& --
& 18.52\%
& 35.80\%
& 14.0 \\

& Heatmap
& Week
& --
& 12.76\%
& 25.51\%
& 23.0 \\

& Heatmap + Statistics
& Activity probability
& Circular + Entropy + Peak
& 15.23\%
& 27.98\%
& 29.0 \\

\cmidrule(lr){2-7}

& \textit{Random baseline}
& --
& --
& 0.41\%
& 2.06\%
& 122.0 \\

\midrule

\multirow{6}{*}{\parbox[t]{0.13\textwidth}{Adolescents Cohort (G)\\($n=262$)}}
& App + Heatmap + Statistics + Transition
& Activity probability
& Circular
& 20.99\%
& 39.31\%
& 11.0 \\

& App + Heatmap + Transition
& Activity probability
& --
& \textbf{21.76\%}
& \textbf{41.98\%}
& \textbf{8.0} \\

& Heatmap
& Activity probability
& --
& 13.74\%
& 29.39\%
& 16.5 \\

& Heatmap
& Week
& --
& 6.87\%
& 17.94\%
& 40.0 \\

& Heatmap + Statistics
& Activity probability
& Circular + Entropy + Peak
& 12.60\%
& 21.37\%
& 36.0 \\

\cmidrule(lr){2-7}

& \textit{Random baseline}
& --
& --
& 0.38\%
& 1.91\%
& 131.5 \\

\midrule

\multirow{6}{*}{\parbox[t]{0.13\textwidth}{Alternative Cohort\\($n=1467$)}}
& App + Heatmap + Statistics + Transition
& Activity probability
& Circular
& 12.75\%
& 23.79\%
& 51.0 \\

& App + Heatmap + Transition
& Activity probability
& --
& \textbf{15.27\%}
& \textbf{29.93\%}
& \textbf{30.0} \\

& Heatmap
& Activity probability
& --
& 8.32\%
& 17.86\%
& 68.0 \\

& Heatmap
& Week
& --
& 7.77\%
& 18.00\%
& 74.0 \\

& Heatmap + Statistics
& Activity probability
& Circular + Entropy + Peak
& 6.82\%
& 14.52\%
& 154.0 \\

\cmidrule(lr){2-7}

& \textit{Random baseline}
& --
& --
& 0.07\%
& 0.34\%
& 734.0 \\

\bottomrule
\end{tabularx}
\end{table*}

\section{Discussion}
\label{sec:discussion}

\begin{table*}[t]
\caption{Processing decisions between event trace and reported feature.
For each step: whether prior screen-time logging studies report the decision, what we observe in our data, and the minimum reporting we propose. Rows marked $\ast$ are quantified in this paper; the remainder are documented as unresolved.}
\label{tab:insights}
\small
\begin{tabular}{@{}p{2.6cm}p{3.4cm}p{4.6cm}p{4.2cm}@{}}
\toprule
Processing step & Reported in prior work & Observation in our data & Minimum reporting \\
\midrule

Threshold derivation$^\ast$ &
Method named (Jenks, $k$-means), derivation of the numeric
boundary not reported~\cite{church_understanding_2015,
morrison_large-scale_2018} &
Jenks and $k$-means yield different thresholds on the same data, and
therefore a different micro-usage classification &
Algorithm, how the boundary is derived from the clusters, and the
resulting numeric value \\
\addlinespace

Application grouping$^\ast$ &
Store taxonomies used without justification; Play Store categories
exclude YouTube~\cite{hamilton_improving_2025} &
Aggregates shift with which applications dominate a sample even when the
category is held fixed; Instagram alone yields markedly different values
than the full set &
Full list of monitored applications, how the set was determined, and
per-application event counts \\
\addlinespace

Temporal binning &
Bin size stated, sensitivity to it acknowledged but not
tested~\cite{hamilton_improving_2025} &
Not varied systematically here; five-minute bins for temporal patterns,
hourly and daily for aggregates &
Bin size, and how sessions crossing a boundary are attributed \\
\addlinespace

Session construction &
Definition of a session given; pairing of opening and closing events not
described &
Only one cohort records closing events, so durations are unavailable
elsewhere; overlapping sessions have no agreed treatment &
Event types recorded, pairing rule, and handling of unmatched or
overlapping events \\
\addlinespace

Missing data &
Data-yield threshold reported for sensing frameworks; authors note it
introduces bias~\cite{hamilton_improving_2025} &
No independent yield signal is available for automation-driven logging,
so absence of events is not separable from absence of logging &
Whether a yield signal exists, and how empty intervals are coded \\
\addlinespace

Participant exclusion &
Sample sizes given; exclusion rules rarely stated &
Activity criteria materially change the analysed sample &
Criteria, counts excluded at each step, and resulting attrition \\
\addlinespace

Device platform &
iOS users excluded from app-level sensing~\cite{hamilton_improving_2025} &
A single automation-driven mechanism records comparable events on both
platforms &
Platforms covered and the collection mechanism used \\
\addlinespace

Time reference &
Rarely stated &
Timezone information is absent from one cohort, weakening its observed
day--night contrast &
Timezone of timestamps and whether it is participant-local \\

\bottomrule
\end{tabular}
\end{table*}

\subsection{From \textit{screen time }to interaction structure}
Research on smartphone use has largely represented a participant by a scalar: minutes of social media per day. The analyses reported here suggest a person is better characterised by the temporal organisation of their interactions when use occurs, across which applications, in what sequence, and how stably over time. Each of the following observations is a consequence of that shift, and none is recoverable from a daily total.
The weekly activity patterns differ and have commonalities across cohorts in ways that daily or weekly totals do not capture in detail. Danish adolescent activity is organised around institutional time  a sharp onset near school start, troughs during lesson blocks, peaks at recess and midday while the German and in-the-wild cohorts show a comparatively unstructured distribution across the afternoon and evening. Two populations with similar aggregate use would be indistinguishable under a totals-based comparison, yet their behaviour is differently organised in time. Within the Danish cohort, the same holds for the effect of the intervention: mean session duration rose from 4.03 to 4.59 minutes between baseline and intervention while per-day use fell, indicating that reduced use was accompanied by consolidation into fewer, longer sessions rather than uniform shrinkage. This is consistent with the mechanism reported by \citet{hansen_disrupting_2026} and is invisible in a duration total. In a nutshell: Aggregate duration is an impoverished representation of digital interaction.

\subsection{The User Is Not a Scalar}
A finding that recurs across the datasets is that aggregate usage statistics are sensitive to which applications dominate a sample. Applications differ systematically in their characteristic session profile: in the Danish baseline data, session counts are dominated by three applications (Snapchat, TikTok, Instagram), whereas Twitch contributes few sessions but a higher proportion of non-micro-usage. A cohort in which a high-frequency, short-session application is prevalent will show elevated opening frequency and reduced mean duration relative to a cohort in which a low-frequency, long-session application dominates (without any difference in underlying behaviour). Because application prevalence varies with age group, region, and recruitment channel, this operates as an unmeasured confound in any comparison across cohorts, including comparisons drawn in the existing literature. Reporting the application composition of a sample should therefore be treated as a minimum requirement rather than a descriptive courtesy. \citet{hamilton_improving_2025} demonstrate the same sensitivity at the level of category definitions; the present results indicate that it persists even when the category is held fixed and only the distribution within it varies .


\subsection{Undocumented Thresholds}
The micro-usage threshold illustrates how far this extends into analysis. \citet{ferreira_contextual_2014} locate the boundary between brief and sustained sessions using Jenks optimization, while \citet{church_understanding_2015} and \citet{morrison_large-scale_2018} use k-means with two clusters. Neither of the latter reports precisely how the numeric boundary was derived from the resulting clusters. Applying all three derivations to the same data produces different thresholds, and therefore a different classification of sessions. The disagreement is modest in absolute terms but it is not recoverable from the published descriptions, which means that a reader cannot determine whether a difference between two studies reflects behaviour or bookkeeping. Since the same ambiguity attaches to session matching, exclusion criteria, and the handling of overlapping sessions, we argue that comparability in this literature now depends less on shared instruments than on shared reporting of the steps between the trace and the feature \cite{ferreira_contextual_2014,morrison_large-scale_2018,church_understanding_2015}.

\subsection{Instrumentation Across Platforms}
The automation-driven logging used here records opening and closing events through the same mechanism on iOS, without requiring persistent background activity or jail braking \citet{morrison_large-scale_2018, kim_real_2019} the device. This addresses a constraint that has shaped the field: \citet{hamilton_improving_2025} excluded iPhone users because iOS does not permit external sensing applications to collect app usage passively, and SensorKit relaxes this only under Apple approval and at category rather than application granularity. Alternative routes exist, including Apple's Screen Time API, yet the literature continues to treat iOS app-level data as largely unobtainable. The practical consequence is that cohorts are split by device platform for reasons unrelated to the research question. The used data collection method works the same on Android and iOS, despite not being demonstrated on Android devices. Demonstrating that a single mechanism can serve both platforms removes one source of incomparability, although it does so at the cost of requiring participants to install and configure a third-party application.
In our view, a key strength of this approach is the transparency afforded by self-installation. Participants install the data-collection automations themselves for each monitored app, making it clear what data are being collected and avoiding reliance on jailbreaking or opaque, pre-granted, blanket collection permissions.

\subsection{Re-identification and the Governance of Usage Data}
The results indicate that app opening traces carry enough individual signal to support re-identification. Specifically, simply temporal traces of activity without further information already allow re-identifying a significant proportion of users, without unique app sets available such in work by \citet{welke_differentiating_2016} and \citet{tu_your_2018}. Temporal activity seems to be individual and mimicking average behavior of others would require knowing their behavior and adjusting everyday usage accordingly\cite{welke_differentiating_2016}. Disappearing into anonymity is thus difficult.
This has a direct bearing on how such data are currently governed. Research ethics review frequently classifies app opening data as low risk on the grounds that no content is captured, and the study protocols underlying these datasets were approved on that basis. This is, for example, the case for the Danish adolescents dataset in this study. The present findings suggest that assessment deserves revisiting. Identity is not the only inference at stake: which applications a person opens may itself disclose sensitive attributes, in the way that \citet{kosinski_private_2013} showed for social media likes. As an example, Grindr ranks seventeenth by opening attempts in the in-the-wild data (N = 22,334), and its presence in a trace is plausibly informative about a user's sexuality. Traces of this kind are routinely collected by third-party applications operating outside a research context and without equivalent oversight. We therefore argue that app usage traces should be treated as potentially identifying and potentially sensitive by default, and that review of studies collecting them (particularly from minors) should assess re-identification risk explicitly rather than inferring low risk from the absence of content. 

\subsection{Towards Reproducible Metrics}
A shared, documented pipeline would address much of the difficulty described above. \textit{RAPIDS} is the closest existing candidate, providing frequency, timing, diversity, episode occurrence, and duration features for application foreground activity, but it supports Android only for app foreground activity \cite{vega_reproducible_2021}. Given that iOS app-level data can be collected, extending such a pipeline to iOS (together with agreed conventions for session matching, exclusion, and the treatment of erroneous or overlapping sessions) would do more for cross-study comparability than further instrument development.

\subsection{Implications for Sensing Research}
The measurement considerations discussed above apply beyond the study of problematic use. Digital phenotyping systems, adaptive interfaces, personal informatics tools, behavioural dashboards, and context-aware applications all derive features from the same class of event traces, and all inherit the same dependence on binning, session matching, application grouping, and threshold derivation or use potentially non-reproducible system APIs outside of researcher control. Where these systems report behaviour back to a user or act on it automatically, undocumented processing choices become design decisions with direct consequences for what the user is told about themselves.








\section{Limitations}
\label{sec:limitations}









\subsection{Sampling and Self-Selection}
\label{subsec:lim-sampling}
 
Three of the four datasets originate from a commercial self-nudging application whose user base consists largely of individuals already motivated to manage their smartphone use consciously. The German and in-the-wild cohorts, recruited in-app, are therefore not representative of smartphone users in general. The Danish cohort was recruited through an online panel and is not subject to this particular bias, but remains confined to a single country, a narrow age band, and iOS devices, and skews female (67.49\%). Danish adolescents cannot be assumed to represent adolescents elsewhere: school schedules, device ownership, and platform popularity all vary by country, and each of these shapes the temporal patterns reported here. Cohort sizes differ by an order of magnitude, which affects the stability of the reported statistics and the comparability of distributional estimates across datasets. Demographic information is unavailable for the German and in-the-wild cohorts.
 
\subsection{What an Opening Event Represents}
\label{subsec:lim-event-semantics}
 
An opening event records that an application was brought to the foreground at a given time. It does not record what occurred inside the application. Posting, messaging, passive scrolling, and checking a notification are indistinguishable in these traces, as are opening an application deliberately and opening it by mistake. The content encountered, the people interacted with, and the reason for opening are all unobserved. This matters because the distinction repeatedly identified as consequential in the psychological literature --- that effects depend on how a platform is used rather than for how long --- falls precisely within what these traces cannot resolve \cite{valkenburg_social_2022}. The measures reported here therefore describe the structure of engagement with applications, not the character of that engagement.
 
\subsection{Intervention Exposure and Retrospective Data}
\label{subsec:lim-intervention}
 
Data collection was not fully passive. For in-the-wild users the intervention is the application's primary function and was active throughout, so the observed behaviour reflects use under an ongoing intervention rather than unmodified use. Recruited cohorts used a non-intervening version of the application; the Danish two-week baseline is consequently the only period in these data during which behaviour was recorded without an active intervention. Parts of the in-the-wild data are retrospective. With the exception of that baseline, this work re-analyses data originally collected in intervention studies, and effects of those interventions cannot be fully separated from the behaviour being described.
 
\subsection{Reactivity}
\label{subsec:lim-reactivity}
 
Participants were aware that their application use was being recorded, and awareness of monitoring may itself alter the behaviour observed. This applies to the Danish baseline period despite the absence of an active intervention, since installing and configuring the logging application makes the observation salient. \citet{hamilton_improving_2025}  report that participants largely forgot their sensing application was installed, which suggests reactivity may attenuate over time, but the present data offer no means of estimating its magnitude .
 
\subsection{Application Selection and Device Coverage}
\label{subsec:lim-app-selection}
 
The Danish cohort monitored a fixed set of ten social media applications. Activity in any application outside that set is invisible in those traces. In the in-the-wild cohort the selection is made by users rather than by design, and monitoring more than one application requires a paid subscription. The resulting application distribution therefore reflects user choice and subscription status rather than usage prevalence, and certain applications are likely overrepresented. This selection has an interpretable upside: users plausibly nominate the applications they themselves experience as most problematic, so the traces may capture self-identified problem use rather than use in general. It nonetheless means the monitored set is neither known in advance nor consistent across users,vwhich limits comparison with cohorts where the application set was fixed.
 
Only activity on the participant's primary smartphone is recorded. Use of the same applications on secondary phones, tablets, laptops, or gaming consoles remains untracked , so the traces understate total engagement by an unknown margin \cite{k_kaye_conceptual_2020}.
 
\subsection{Instrumentation and Measurement Artefacts}
\label{subsec:lim-instrumentation}
 
Only the Danish cohort recorded both opening and closing events and therefore yields session durations; the German and in-the-wild cohorts record opening attempts together with their outcome, supporting analyses of frequency and abandonment but not of duration. The micro-usage analysis is restricted accordingly. Timezone information is absent from the in-the-wild raw data, so timestamps are reported in UTC and the weaker day--night contrast in that cohort is at least partly an aggregation artefact rather than a behavioural difference.
 
Devices supporting simultaneous display of multiple applications present a further difficulty: concurrently active applications would produce overlapping sessions under the present logging approach, inflating recorded screen time through double counting. This is one of several reasons to treat session duration as an insufficient measure on its own. Operating system behaviour and device configuration can additionally suppress events, and such losses are not distinguishable from genuine non-use.
 
\subsection{Preprocessing}
\label{subsec:lim-preprocessing}
 
There are no agreed conventions in this literature for the preprocessing steps that precede analysis, and the choices made here are defensible rather than standard. This applies to the exclusion of users and traces (whether by minimum or maximum event count, by required observation duration, or by activity criteria) to the matching of opening and closing events into sessions, and to the handling of apparently erroneous records such as overlapping sessions. Each of these alters the resulting metrics. Free-living data additionally contain noise, missing values, and uncontrolled variation that cannot be resolved post hoc: intervals without recorded events cannot be distinguished from intervals in which
logging did not run, as no independent data-yield signal is available.
 
\subsection{Privacy}
\label{subsec:lim-privacy}
The traces analysed here are individually distinctive, and the inferences they support extend beyond identity to potentially sensitive attributes. This is discussed and in the ethics and privacy statement. It is noted here because it constrains how such datasets can responsibly be shared, and the data underlying this work cannot be released in raw form. 
 
\subsection{Scope of Claims}
\label{subsec:lim-scope}
 
No mental health or well-being outcomes are analysed here, and no causal claims about the relationship between usage patterns and psychological outcomes are made or implied. The Danish baseline and intervention periods differ in length (two and four weeks respectively), so raw totals across periods are not directly comparable and per-day rates are reported where that comparison is drawn. Comparison with \citet{hamilton_improving_2025} holds at the level of measures rather than measurement: their estimates derive from hourly-binned foreground sensing across all applications on Android devices, whereas ours derive from session-level events for a fixed or user-selected application set, predominantly on iOS. Correspondence should therefore be read as evidence that the shape of the findings is robust to instrumentation, not as agreement between comparable quantities.

\section{Conclusion}
\label{sec:conclusion}
This work demonstrates that smartphone app activity logs capture substantially richer behavioral information than aggregate screen-time measures alone. Across three cohorts, we observed temporal routines, application-specific patterns, transitions between apps, and individual behavioral signatures that a single measure of daily duration would obscure. These patterns also reveal how changes in smartphone use can occur without corresponding changes in total time (e.g. through shifts from fewer, longer sessions to more frequent, shorter interactions).

Our results also highlight that the insights obtained from activity logs depend strongly on methodological choices. Differences in event definitions, session construction, temporal aggregation, application selection, and thresholds for micro-usage can change the resulting behavioral measures and limit comparability across studies. We therefore encourage HCI researchers to report these transformations explicitly and to treat the event-to-feature pipeline as part of the research contribution rather than an implementation detail.

At the same time, greater behavioral resolution introduces greater privacy risk. The ability to distinguish individuals from seemingly simple app-opening traces demonstrates that such data can constitute a sensitive behavioral fingerprint, particularly for vulnerable populations such as adolescents. Future HCI work should therefore pursue richer sensing alongside privacy-aware data minimization, transparent reporting, and reproducible cross-platform methodologies.

More broadly, our findings suggest a shift from asking \textit{how much} people use apps toward understanding \textit{when, how, and in what patterns} that use occurs.

\begin{acks}
DG’s work is funded by the Huo Family Foundation and Stanford’s Center for Digital Health. DG has ongoing research projects in collaboration with the one sec app. PS’s work is funded by Stanford’s Center for Digital Health.
\end{acks}

\section*{Ethics and Privacy Statement}

This work characterizes behavioral patterns in app-opening traces from over 1,900 adolescents and adults, with implications for how such data are governed in research and industry practice. The primary benefit is methodological: these findings demonstrate that fine-grained temporal activity patterns (beyond total screen time) are critical for understanding smartphone interaction, with potential to improve the design of interventions for problematic use.
However, we identify a significant governance concern. All four original studies underlying this secondary analysis obtained appropriate ethics  approval from institutional review boards or equivalent governance bodies (see Methods).
All original studies underlying this secondary analysis obtained appropriate ethics approval from institutional review boards or equivalent governance bodies (see Methods). These approvals were grounded in the then-prevailing assumption that app-opening data (lacking in-app content) posed minimal privacy risk.

\section*{Author Contributions}
\noindent\textbf{Conceptualization:} OS, PG, DG.\\
\noindent\textbf{Methodology:} OS, PG.\\
\noindent\textbf{Investigation:} FR, DW, DG.\\
\noindent\textbf{Data curation:} OS.\\
\noindent\textbf{Formal analysis:} OS.\\
\noindent\textbf{Software:} OS.\\
\noindent\textbf{Validation:} OS, PG.\\
\noindent\textbf{Visualization:} OS.\\
\noindent\textbf{Writing - original draft:} OS, PG.\\
\noindent\textbf{Writing - review and editing:} OS, PG, FR, DW, AS, PS, DG.\\
\noindent\textbf{Project administration:} PG, DG.\\
\noindent\textbf{Supervision:} AS, PS, DG.



\end{document}